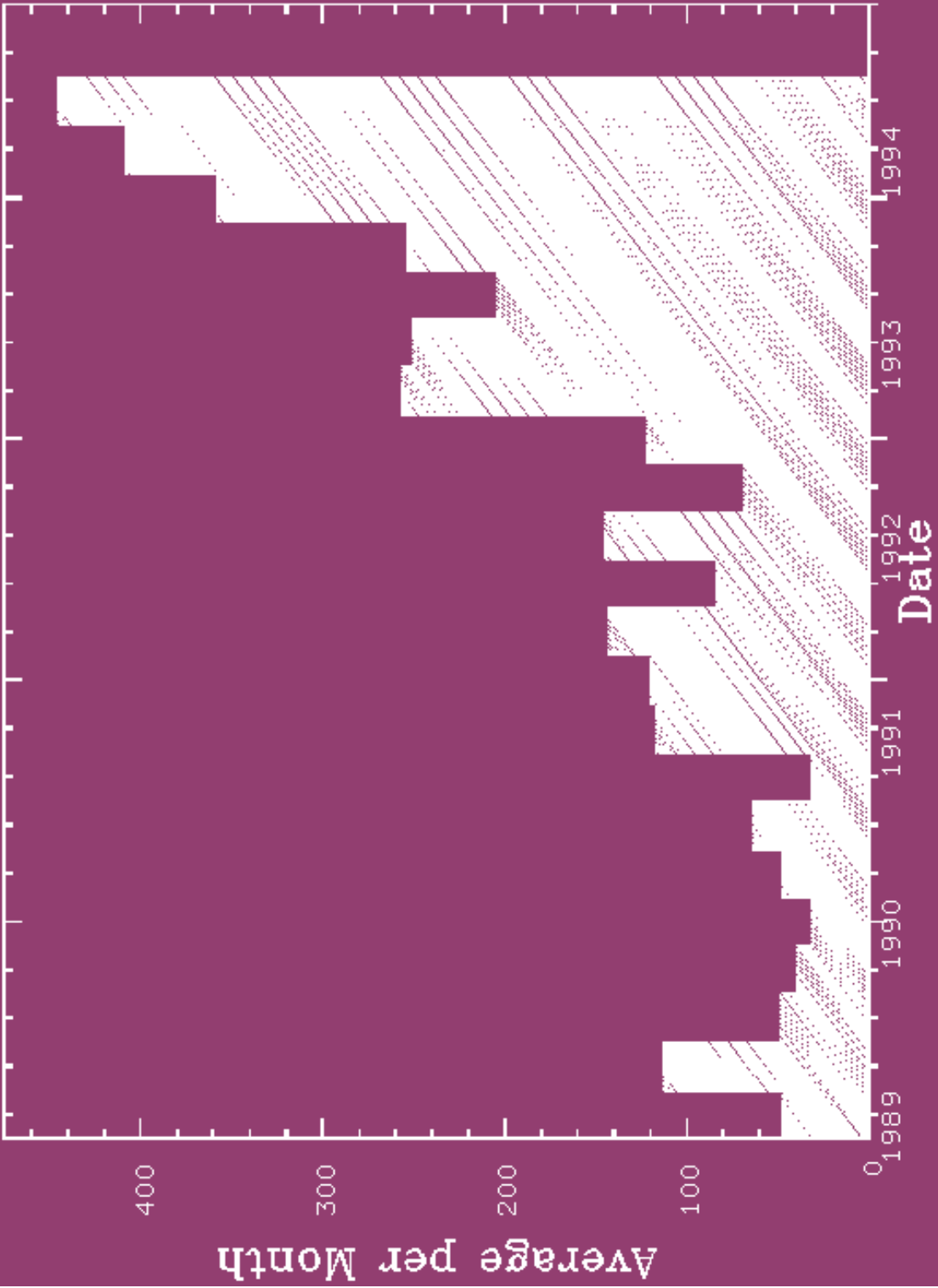

# The EINSTEIN On-Line Service

D. E. Harris and C. S. Grant

*Center for Astrophysics, 60 Garden St., Cambridge, MA 02138*

H. Andernach

*Observatoire de Lyon, F-69561 Saint-Genis-Laval Cedex, France*

**Abstract.** The Einstein On-Line Service (EOLS) is a simple menu-driven system which provides an intuitive method of querying over one hundred database catalogs. In addition, the EOLS contains over 30 CDROMs of images from the Einstein X-ray Observatory which are available for downloading. The EOLS provides all of our databases to the NASA Astrophysics Data System (ADS) and our documents which describe each table are written in the ADS format. In conjunction with the IAU working group on Radioastronomical Databases, the EOLS serves as an experimental platform for on-line access to radio source catalogs. The number of entries in these catalogs exceeds half a million.

## 1. INTRODUCTION

In January 1989, SAO established an on-line service to help astronomers prepare ROSAT proposals by providing access to the preliminary source list from the "Einstein Observatory Catalog of IPC X-ray Sources". In the intervening years, we have updated the source list, added to the documentation, included many more Einstein databases as well as a number of tables from other wavebands, provided access to images for downloading from all of the Einstein CDROMs, and installed new software for more sophisticated filtering and retrieval. Although we have improved the functionality and made significant additions to the databases, we still maintain a simple menu interface accessible from any type of terminal. An instruction manual does not exist: the on-line help facility provides the necessary information.

## 2. DATABASE RETRIEVAL OPTIONS

The EOLS provides three options for data retrieval. The DATABASE QUERY is a menu-driven system which allows users to build their own query and select which output columns should be retrieved. The QUICK QUERY is a canned output format which requires only one specification of selection criteria, and returns up to 79 characters per row (preselected by us). The MULTIPLE QUICK QUERY is a positional search on several tables. We have endeavored to make the selection process simple and the easiest option is to select one or more affinity groups. However, the user can specify particular tables from different groups by choosing the "Custom" option or choose to select all tables.

Here are 6 of the 30 rows retrieved from 105 catalogs with a single query:

```
--------------------------- eos_source ------------------------------
| SEQ |FLD|CAT |    RA     |   DEC    |+/-|COR C/S|       |    |      |    |   |
|  #  | # | #  | h  m   s  | d  '  '' |asc|cts/sec|  +/-  | S/N|SIZCOR|RECO| ID|
|-----|---|----|-----------|----------|---|-------|-------|----|------|----|---|
| 3042|  1|4594|22 17 41.3 |+08 44 55 | 36|2.0e-02|4.4e-03| 4.4|  1.1 |   0| Q |
---------------------------------------------------------------------
Total: 1 rows retrieved.

--------------------------- qso_veron85 -----------------------------
|            |    RA     |    DEC   |     |     |radio  | fdr6   | fdr11  |
|        name| h  m   s  | d  '  '' |   z | vmag|source?| (cgs)  | (cgs)  |
|------------|-----------|----------|-----|-----|-------|--------|--------|
|            |22 17 39.4 |+08 44 56 |0.228|17.60|   yes |7.50e-25|0.00e+00|
|            |22 17 42.5 |+08 45 24 |0.623|18.60|   yes |7.00e-25|0.00e+00|
---------------------------------------------------------------------
Total: 2 rows retrieved.

--------------------------- rad_6cm_bwe91 ---------------------------
|          |    RA     |    DEC   |S(4.85GHz)|Extension|Spectral|Separation|
|   Name   | h  m   s  | d  '  '' |   (mJy)  |   Flag  |  Index |   Flag   |
|----------|-----------|----------|----------|---------|--------|----------|
|2217+0844 |22 17 42.0 |+08 44 44 |     121  |         |   -0.6 |          |
---------------------------------------------------------------------
Total: 1 rows retrieved.

--------------------------- rad_gb87 --------------------------------
|    RA     |RA+/-|   DEC    |DEC+/-|S(4.85)|S +/- |EXT|WARN|CONF|FWHM|FWHM| PA|
| h  m   s  | sec | d  '  '' | arcs | (mJy) |(mJy) |FLG|FLAG|FLAG|MAJ |MIN |   |
|-----------|-----|----------|------|-------|------|---|----|----|----|----|---|
|22 17 41.7 | 0.8 |+08 44 45 |  13  |  125  |  18  |   |    |    |1.20|1.01|-71|
---------------------------------------------------------------------
Total: 1 rows retrieved.

--------------------------- rad_pks90 -------------------------------
| +/- |   |    |        |fdr1410 |fdr2700 |fdr5000 |    RA     |    DEC   |       |
| arcs| id| mag|       z|  (Jy)  |  (Jy)  |  (Jy)  | h  m   s  | d  '  '' | alias |
|-----|---|----|--------|--------|--------|--------|-----------|----------|-------|
|   20| Xs|    |        |  0.230 |        |        |22 17 39.7 |+08 44 58 |4C08.66|
---------------------------------------------------------------------
Total: 1 rows retrieved.
```

## 3. DOWNLOADING X-RAY IMAGES

All of our Einstein CDROMs are permanently available for downloading images. They may be obtained with ftp or with the VMS "copy" command. As of 30 June 1994, we have:

The 2E Catalog of IPC X-ray Sources: FITS smoothed arrays
The Database of HRI X-ray Images: FITS photon arrays
The IPC Slew Survey: 1 Jan 1991
The HRI Images in Event List Format: FITS binary tables
The IPC Images in Event List Format: FITS binary tables
The IPC Unscreened Data Archive: FITS binary tables

## 4. OTHER FEATURES

- Low maintenance operation.

- The ADS connection – The EOLS provides all of our databases to the NASA Astrophysics Data System (ADS) and our documents which describe each table are written in the ADS format.

- DOCUMENTATION – Documentation is available for each database table and provides extensive detail in some areas (e.g. Vol 1 of the IPC catalog), archival material such as the Einstein Revised User's Manual, and "Help" files which provide information on system operations.

- COMMUNICATION – This section of the EOLS features a bulletin board (includes ROSAT news) and an email facility for sending yourself documents and the results of database searches.

- TOOLS – Precession of coordinates and column densities of galactic HI (north of DEC = -40 degrees) are available.

- ACCESS – There are four methods to log onto the EOLS:
    1. internet – telnet einline.harvard.edu
    2. decnet – set host 6714
    3. modem – dial (617) 495-7047 or 495-7048
    4. mosaic – http://hea-www.harvard.edu/einline/einline.html

Once successful, the login name is "einline", no password required.
NB: Since UNIX is case sensitive, make sure you use lower case when logging in!

## 5. THE RADIO SOURCE CATALOG INITIATIVE

For many years we have noticed the difficulty of finding radio source data in public archives or on-line databases. In 1991, the IAU Working Group on *Radioastronomical Databases* was formed with one of us (H.A.) as the chair. By now the WG has secured some 130 catalogs of radio sources with altogether ~560,000 entries. Although the working group recognized that a long term solution to this problem should be provided by an on-going commitment from the National Radio Astronomy Observatory (NRAO) or the Strasbourg Astronomical Data Center (CDS), the EOLS volunteered to serve as an experimental platform for access to the catalog collection. This "provisional" solution is now by far the most complete on-line facility for radio source data, offering access to 63 radio catalogs with 520,000 searchable entries. Contributions of new data tables are welcome, but can only be incorporated into the system if authors provide proper documentation and formatted ASCII tables.

## 6. STATISTICS OF USAGE

The histogram shows the average number of logins per month from outside the Center for Astrophysics from 1989 to mid 1994.

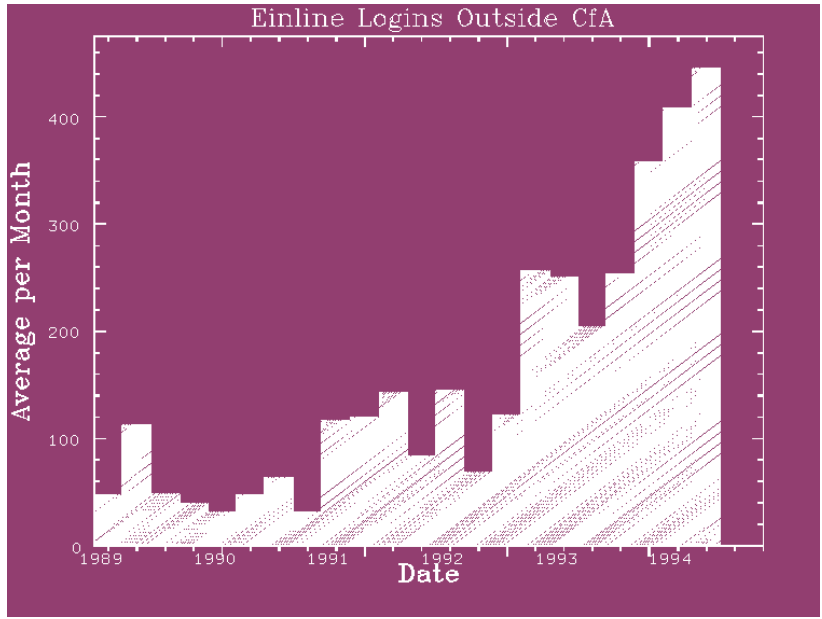

Figure 1.    Logins per month from outside CfA: 1989–1994